\documentclass[useAMS,usenatbib, usegraphicx, usedcolumn]{mn2e}

\title[Effect of Be Disk Evolution on Global One-Armed Oscillations]
  {Effect of Be Disk Evolution on Global One-Armed Oscillations}
\author[F.Oktariani et al.]
  {F.~Oktariani,$^1$ \thanks{Email: f.oktariani@math.itb.ac.id}
  A.T.~Okazaki,$^2$ C.~Kunjaya,$^3$ Aprilia$^3$\\
  $^1$Combinatorial Mathematics Research Group, FMIPA, Institut Teknologi Bandung, Ganesha 10, Bandung 40132, Indonesia \\
	$^2$Faculty of Engineering, Hokkai-Gakuen University, Toyohira-ku, Sapporo 062-8605, Japan\\
  $^3$Department of Astronomy, FMIPA, Institut Teknologi Bandung, Ganesha 10, Bandung 40132, Indonesia}
\date{Released 2015 Xxxxx XX}

\pagerange{\pageref{firstpage}--\pageref{lastpage}} \pubyear{2015}

\def\LaTeX{L\kern-.36em\raise.3ex\hbox{a}\kern-.15em
    T\kern-.1667em\lower.7ex\hbox{E}\kern-.125emX}

\begin{document}

\label{firstpage}

\maketitle

\begin{abstract}
 We study the effect of density distribution evolution on the global one-armed oscillation modes in low viscosity disks around isolated and binary Be stars. Observations show that some Be stars exhibit evidence of formation and dissipation of the equatorial disk. In this paper, we first calculate the density evolution in disks around isolated Be stars. To model the formation stage of the disk, we inject mass at a radius just outside the star at a constant rate for $30-50$ years. As the disk develops, the density distribution approaches the form of the steady disk solution. Then, we turn off the mass injection to model the disk dissipation stage. The innermost part of the disk starts accretion, and a gap forms between the star and the disk. Next, we calculate the one-armed modes at several epochs. We neglect the effect of viscosity because the time-scale of oscillations is much shorter than the disk evolution time-scale for low viscosity. In the disk formation stage, the eigenfrequency increases with time toward the value for the steady state disk. On the other hand, one-armed eigenmodes in dissipating Be disks have  significantly higher eigenfrequencies and narrower propagation regions. 
Observationally, such a change of mode characteristics can be taken as an evidence for gap opening around the star. In binary Be stars, the characteristics of the disk evolution and the eigenmodes are qualitatively the same as in isolated Be stars, but quantitatively they have shorter evolution time-scales and higher eigenfrequencies, which is in agreement with the observed trend.
\end{abstract}


\section{Introduction}
\label{sec:intro}
Be stars are non-supergiant B-type stars whose spectra have, or have
had at some time, one or more Balmer lines in emission. The optical emission lines arise from an equatorial disk around the star. It is a geometrically thin, high-density plasma that rotates at a  nearly Keplerian speed (e.g., \citealt{Por03}). The most promising model of the equatorial disk is the viscous decretion model (\citealt{Lee91}), where the material ejected from the central star forms a disk, drifting outward because of the angular momentum transfer by viscosity.

Many Be stars show variations in their spectra over years to decades. One of them is called long-term V/R
variations, which are variations of the ratio of relative intensity of violet (V) and red (R) peaks of a double-peaked
emission line profile. It is widely accepted that the long-term V/R variations are attributed to global one-armed (i.e., the azimuthal wave number $m=1$) oscillations in the Be disk (e.g., \citealt{Por03}). It is based on the fact that in a nearly Keplerian disk a one-armed mode shows up as a very slowly revolving perturbation pattern (e.g., \citealt{Kat83}). Early studies of the one-armed oscillation model (\citeauthor{Oka91} \citeyear{Oka91}, \citeyear{Oka97}; \citealt{Pap92, Sav93, FH06}) were lacking the mechanism for confining the modes to the inner part of the disks. Recent development of the models by \citet{Ogi08} solved this problem by including three-dimensional effect that contributes to the mode confinement. All of the studies mentioned before were for isolated Be stars. But, there are also many Be stars found in a binary system where the presence of the companion star is likely to change some features of the global one-armed modes. \citet{Oct09} studied global one-armed oscillations in binary Be star systems taking into account the effects of tidal interaction with the companion star.

Most of the above studies assumed a power-law density distribution in the equatorial disk.
This had first been done just for simplicity, where the density slope was just a parameter.
After the scenario has been widely accepted that Be disks are formed by viscous diffusion
of material ejected from the central star \citep{Lee91},
the $\Sigma \propto r^{-2}$ distribution has often been adopted, which corresponds to
the density distribution of the steady decretion disk solution (\citealt{Por99}; \citealt{Oka01};
see also \citealt{Krt11}).

However, observational results show that some Be stars exhibit
evidence of formation and dissipation of the equatorial disk (\citealt{Riv01};\citealt{WI10}), during which the density distribution in the disk is likely far from the usually-assumed, power-law distribution. For example, in the disk dissipation stage, the disk is expected to be lost from the innermost part, causing a gap between the disk and the star. Such a long term change in the density distribution is likely to affect the one-armed oscillation modes. 

The studies of the effect of disk density evolution to the properties of the Be stars, such as photometric and polarimetric properties, have been done in recent years. For example,  \citet{HA12} investigated the evolution of continuum photometric observables caused by the evolution of density in Be disk. The study includes different density phases, such as the disk build-up phase and disk dissipation phase. In general, the results from their model are in agreement with observations. In our model, we use nearly the same method as \citet{HA12} in calculating the evolution of the disk density. \citet{HA14} also investigated the evolution of continuum polarimetric observables for different dynamical scenarios, where they gave theoretical prediction of the polarimetric properties in a star with an evolving disk. \citet{HAL13} studied the intrinsic continuum linear polarisation of Be star during the formation and dissipation of the circumstellar disk. They show that the polarimetric properties can be used to determine the mechanism for mass decretion from the central star. 

In this paper we consider the effects of the Be-disk evolution, in order to study the global one-armed oscillations with a more realistic disk model. The structure of the paper is as follows. In section ~\ref{sec:evol}, we discuss the disk evolution around single and binary Be stars.
In section ~\ref{sec:oscill}, we calculate linear, one-armed perturbations for the disk structure
obtained in the previous section. Sections ~\ref{sec:discussion} and ~\ref{sec:conclusion} are for discussion and conclusions.

\section{Modeling the Formation and Dissipation of Viscous Be Disks}
\label{sec:evol}

Below we calculate the density evolution in a viscous disk around a Be
star. As the central Be star, we take a $\mathrm{B0V}$ star with
$M_{1} = 17.5$ $M_{\odot}$, $R_{1} = 7.4$ $R_{\odot}$, and
$T_{\mathrm{eff}}=30000$ $\mathrm{K}$ \citep{Cox00} {where $M_{1},R_{1}$ and $T_\mathrm{eff}$ is the mass, radius, and effective temperature of the star. For simplicity, we do not include the effect of the rotation on the shape of the star. We assume that the Be disk is geometrically thin, axisymmetric and nearly Keplerian, and is in hydrostatic equilibrium in the vertical direction. For simplicity, we also assume that the disk is isothermal at $0.6$ $T_\mathrm{eff}$ \citep{Car06}, where $T_\mathrm{eff}$ is the effective temperature of the central star. The evolution of such a disk is described by
the following one-dimensional, diffusion-type equation of the surface density $\Sigma$,

\begin{equation}
 \frac{\partial \Sigma}{\partial t} =  \frac{1}{r} \frac{\partial}{\partial r} \left[  \frac {\frac{\partial}{\partial r} (r^{2}\Sigma \alpha c_{s}^{2} ) }{\frac{d}{d r} (r^{2} \Omega ) } \right]
\label{eq:density}
\end{equation}
(see e.g. \citealt{Pri81}; see also \citealt{Oka02} for discussion in the context of Be disks),
where $\alpha$ is the Shakura-Sunyaev viscosity parameter , $c_{s}$ is the isothermal sound speed, and $\Omega$ is the angular frequency of the disk rotation. In this section, we take the angular frequency to be the Keplerian angular frequency $\Omega_{K}$ where $\Omega_{K} = (GM/r^{3})^{1/2}$.\footnote{ We will take account of small deviation of $\Omega$ from the Keplerian angular frequency,
when we consider one-armed oscillation modes in the next section.}

\subsection{Disk Evolution Around Isolated Be Stars}
\label{sec:evol-single}

In order to model the formation stage of a disk around an isolated Be star, we inject mass at a constant rate at a radius just outside the star. The mass injection radius, $r_{\rm inj}$, can be arbitrarily chosen, because it affects little the resulting density distribution as long as $r_{\rm inj}-R_1 \ll R_1$. In this paper we take $r_{\rm inj}=1.008$ $R_{1}$.
At the inner boundary $r = R_{1}$, we adopt the torque-free boundary condition, $\Sigma=0$, so matter will only move outward. Since the density becomes close to zero at large distances, and the region of interest for one-armed oscillations is at most $\la 100$ $R_1$, we adopt that $\Sigma=0$ at $r=1000$ $R_{1}$, as the outer boundary condition.

\begin{figure}
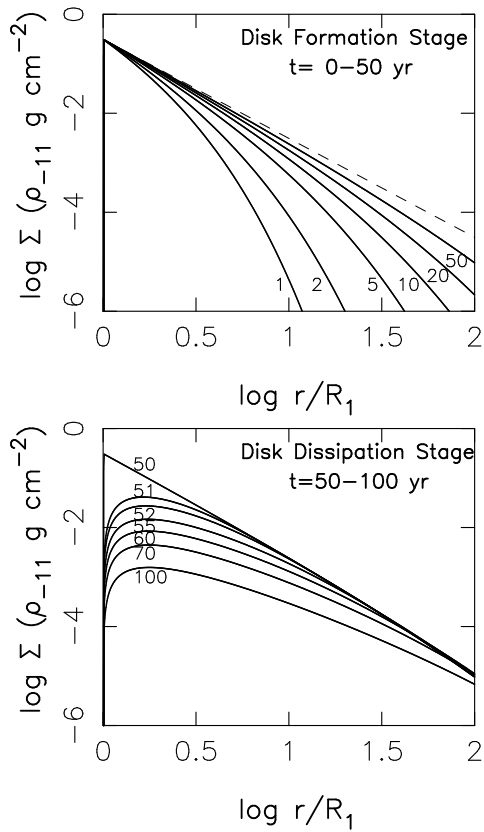

  \begin{center}
	\includegraphics[width=63mm]{ibdmfv.eps}
  \includegraphics[width=63mm]{ibdmdv.eps}		
	\end{center}
	\caption{Evolution of surface density distribution in the disk
          of an isolated Be star. The upper panel is for the formation
          stage, and the lower panel is for the 
          dissipation
          stage. We take $\alpha = 0.1$ as the viscosity
          parameter. The mass injection rate adopted is $1.2 \times
          10^{-9} M_{\odot} \mathrm{yr}^{-1}$, but the resulting
          density distribution is arbitrarily scalable.  The { dashed} line in the upper panel denotes the density distribution with the power-law form, $\Sigma \sim r^{-2}$. Annotated to each line is the elapsed time in years. }
    \label{fig:devos}
\end{figure}

In the upper panel of Fig.~\ref{fig:devos}, we show the evolution of $\Sigma$ over 50 years  in the disk formation stage. In this calculation, we took $\alpha = 0.1$ and  $\dot{M} = 1.2 \times 10^{-9} M_{\odot} \mathrm{yr}^{-1}$, which gives rise to the typical base density of $10^{-11} {\rm g\,cm}^{-3}$. Although we adopted a particular mass injection rate, $\Sigma (r)$ is proportional to the local mass injection/decretion/accretion rate in isothermal disks, so that the resulting density distribution is arbitrarily scalable.

In the early times of the formation stage, the Be disk has steep density distributions: the mass is concentrated in the region near the star. However, by viscous diffusion, the injected mass gradually flows to the outer part. Therefore, the surface density distribution becomes flatter with time. In principle, as $t \to \infty$, it approaches the steady state density distribution $\Sigma \sim r^{-2}$, especially in the inner part of the disk, which is shown by the dashed line in the upper panel of Fig.~\ref{fig:devos}. From the figure, we see that at $t \sim 50\,{\rm yr}$, the surface density in the inner $10$ $R_{1}$ has almost reached this state.

In order to model the disk dissipation, we turned off the mass injection at $t=50\,{\rm yr}$, and calculated the surface density evolution for another $50$ years. The result is shown in the lower panel of Fig.~\ref{fig:devos}. From the figure, we note that the accretion begins as soon as the mass injection stops. The disk is lost from the innermost part, causing a gap between the star and the disk. The gap forms because the viscous timescale, which is proportional to $r^{1/2}$ in isothermal disks, is shorter at smaller radii. Then, the density distribution has a peak near the inner radius.

\subsection{Disk Evolution Around Binary Be Stars}
\label{sec:evol-bin}

\begin{figure}
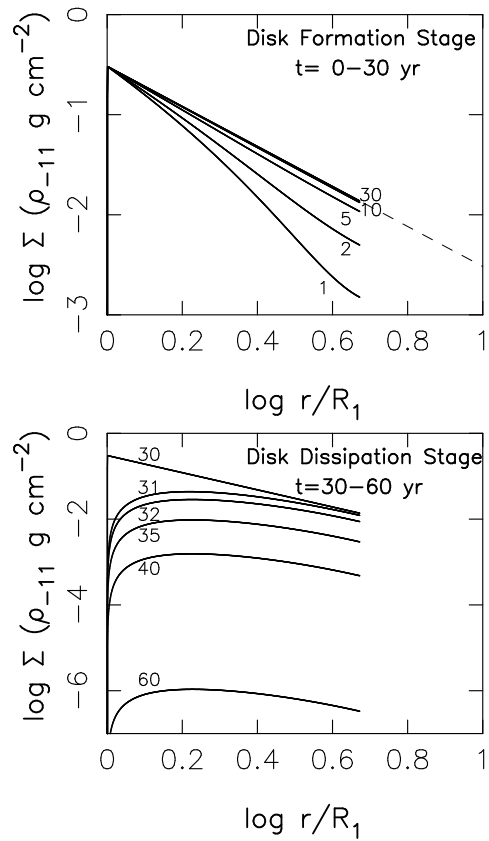

 \begin{center}
	\includegraphics[width=63mm]{bbdmfd1.eps}
	\includegraphics[width=63mm]{bbdmdd1.eps}	
	\end{center}		
	\caption{Evolution of the surface density distribution in the disk of a binary Be star with orbital separation $D = 10$ $R_{1}$ and binary mass ratio $q = 0.1$. The upper panel is for the formation stage, and the lower panel is for the 
	dissipation stage. The { dashed} line in the upper panel shows the density slope for $\Sigma \sim r^{-2}$. Annotated to each line is the elapsed time in years. The viscosity parameter and the mass injection rate are the same as in the isolated Be star cases. }
    \label{fig:devob}
\end{figure}

\begin{figure}
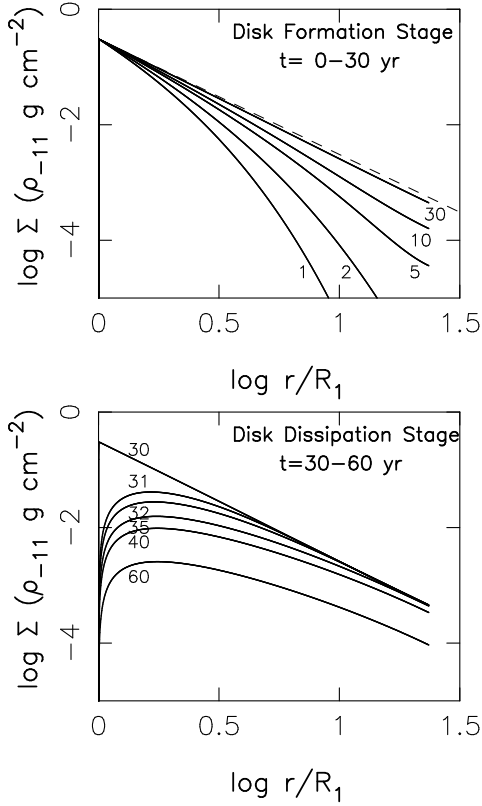

   \begin{center}
	   \includegraphics[width=63mm]{bbdmfd5.eps}
	   \includegraphics[width=63mm]{bbdmdd5.eps}	
	\end{center}		
	  \caption{Same as Fig.~\ref{fig:devob}, but for $D = 50$ $R_{1}$.}
    \label{fig:devob5}
\end{figure}

In binary systems, a circumstellar disk, like a Be disk, is tidally/resonantly truncated at a radius depending on the companion mass and the orbital eccentricity (\citealt{Art94}; see also \citealt{ON01}).
In this section we study the effect of companion on the Be disk evolution, by taking into account
the disk truncation. {As in \citet{Oct09}, we consider, for simplicity, circular binaries with small mass ratios
($M_{2}/M_{1} \la 0.3$), with $M_{2}$ being the mass of the companion. However, there are many interesting Be binaries with non-zero orbital eccentricities and higher mass ratios. It is thus desirable to consider those systems in future studies. 
In a circular binary, a viscous disk is truncated at the tidal radius given by
\begin{equation}
R_\mathrm{tides} \sim 0.9 R_\mathrm{L}
\label{eq:rtides}
\end{equation}
\citep{WK91}. Here $R_\mathrm{L}$ is the Roche lobe radius of the Be star given approximately by
\begin{equation}
R_\mathrm{L} = D \frac{0.49q^{-2/3}}{0.69q^{-2/3} + \ln (1+q^{-1/3})}
\label{eq:roche}
\end{equation}
\citep{Egg83}, where $D$ is the orbital separation and $q = M_{2}/M_{1}$ is the binary mass ratio.
In this paper, we assume that the Be disk is truncated at $r=R_\mathrm{tides}$.

As in the case of isolated Be stars, we model the disk evolution by injecting mass at a constant rate at $r=1.008$ $R_{1}$. Given a shorter viscous time-scale in a truncated disk than in an isolated disk, we run the calculation over 30 years. We then turn off the mass injection and calculate the disk evolution for the next 30 years to study the disk dissipation stage. We adopt the free boundary condition, $\Sigma = 0$, at the inner boundary $r=R_{1}$. As the outer boundary, we take $V_{r} = 0$
to emulate the effect of disk truncation,
where $V_{r}$ is the radial velocity given by
\begin{equation}
V_r = - \frac{\frac{\partial}{\partial r} (\alpha c_{s}^{2} r^{2} \Sigma)}{r \Sigma \frac{\mathrm{d}}{\mathrm{d}r} (r^{2} \Omega)}
\label{eq:vrdisk}
\end{equation}
\citep{Pri81}.

Figures ~\ref{fig:devob} and ~\ref{fig:devob5} show the surface density evolution of the Be disks in binary systems with $D = 10$ $R_{1}$ and $D = 50$ $R_{1}$, respectively. In both figures, we took binary mass ratio $q=0.1$ and the viscosity parameter $\alpha = 0.1$. The upper panel is for the disk formation stage and the lower panel is for the disk dissipation stage.

The initial evolution of the surface density distribution of Be disks in binary systems is similar to that in the isolated Be stars. However, at later times, after the disk material reached
the outer radius, the influence of the tidal truncation is seen in the disk structure, especially in the outermost part of the disk. This results in the difference in the surface density distribution between disks in the binary and isolated Be stars.
With a shorter viscous time-scale and the stagnation of the outflow at the outer radius,
the Be disk in binary systems more rapidly grows and approaches the $\Sigma \sim r^{-2}$ distribution
than in isolated Be stars.

Another notable feature of Be disk evolution in binaries is also related to the disk size. The Be disk evolves more quickly in closer binary systems because of smaller disk sizes.
We can see this in the upper panels of Figs.~\ref{fig:devob} and ~\ref{fig:devob5}.
In the dissipation stage, the Be disk evolution is also quicker in closer binaries.
However, the very rapid decrease in density seen in the lower panel of Figs.~\ref{fig:devob}
has another reason. In the disk dissipation stage, the accretion starts from the innermost region,
and the accreting region becomes larger with time.
The local rate of surface density decrease is given by the difference between the accretion rate at the radius considered and the rate of mass accreting into that radius from outer radii, the latter of which
is lower for smaller disks. This is why the surface density in disks for $D=10$ $R_{1}$
decreases much more rapidly than in disks for $D=50$ $R_{1}$, which decreases more rapidly
than in disks of isolated Be stars.
{It is worth noted that there are other mechanisms that can cause the disk loss in the innermost region such as the ablation of the inner disk by radiation and/or wind of the central star. We further discuss these possibilities in section~\ref{sec:discussion}.}

\section{Global One-Armed Oscillations}
\label{sec:oscill}

In this section, we seek linear $m=1$ eigenmodes for the density distribution
obtained in the previous section. Since the time-scale of oscillations is
much shorter than the time-scale of viscous disk evolution, we neglect the effect
of viscosity in the following analysis.
We also neglect the radial velocity $V_r$ in the unperturbed disk.
For simplicity, we also assume that the oscillations are isothermal.
Then, the basic equations for $m=1$ oscillations to be solved are the same as
in \citet{Oct09}, except that we use the density distribution in forming/dissipating disks
in this paper.

Below we first summarize the unperturbed disk model and basic equations for linear $m=1$ isothermal perturbations (for more details, see \citealt{Oct09}). Then, we present the $m=1$ eigenmodes
in disk formation/dissipation stages for single and binary Be-star cases.

\subsection{Basic Equations}

The unperturbed state of the isothermal Be disk is derived from
the equation of hydrostatic equilibrium in the vertical direction and
the radial component of the equation of motion as
\begin{eqnarray}
\rho &=& \rho_{m}(r) \exp \left( - \frac{z^{2}}{2H^{2}}\right),
\label{eq:rho_z}
\\
\bf{v} &=& [0, r\Omega(r),0],
\label{eq:unperturb-v}
\\
p &=& c_\mathrm{s}^2 \rho,
\label{eq:unperturb-iso}
\end{eqnarray}
and
\begin{equation}
\Omega=  \left[\frac{1}{r}\left(\frac{d\Psi_\mathrm{m}}{dr} + c_{s}^{2}\frac{d\ln\rho_\mathrm{m}}{dr}
\right)\right]^{1/2},
\label{eq:centrifugal}
\end{equation}
where $\rho_{m} (r)$ is the midplane density, and
$H(r) = c_\mathrm{s}/\Omega_\mathrm{K}$ is the scale-height of the disk,
$\Omega(r)$ is the angular frequency of disk rotation,
and $\Psi_\mathrm{m}$ is the potential in the disk midplane.
Here, $\rho_{\rm m}(r)$ is calculated from
the surface density distribution obtained in section~\ref{sec:evol},
using the relation $\Sigma=\sqrt{2\pi}\rho_{\rm m}H$ derived from equation~(\ref{eq:rho_z})
for isothermal disks.
The potential $\Psi_\mathrm{m}$, where the quadrupole contribution due to
the rotational distortion of the central star and the azimuthally averaged tidal potential
are taken into account, is given by
\begin{eqnarray}
\Psi_\mathrm{m} & \simeq & -\frac{GM_{1}}{r} \left[ 1 + k_{2}\left(\frac{\Omega_{1}}{\Omega_\mathrm{c}}
 \right)^{2} \left(\frac{r}{R_{1}}\right)^{-2} \right] \nonumber\\
 && - \frac{GM_{2}}{D}\left[ 1 + \frac{1}{4}{\left(\frac{r}{D}\right)}^2\right],
\label{eq:potential}
\end{eqnarray}
where $k_{2}$ and $\Omega_\mathrm{c}=\sqrt{GM_{1}/R_{1}^{3}}$ are
apsidal motion constant and critical angular rotation speed of the Be
star, respectively. { $M_{2}$ is the mass of the companion star. In
  the case of isolated Be star, we set $M_{2} = 0$.}

On the above unperturbed disk model,
we superpose a linear $m=1$ isothermal perturbation of frequency $\omega$, which varies as
$\exp[i(\phi-\omega t )]$.
The linearized perturbed equations are then obtained as follows:
\begin{equation}
i(\Omega - \omega) v_{r}^{\prime} - 2\Omega v_{\phi}^{\prime} = - \frac{\partial h^{\prime}}{\partial r},
\end{equation}
\begin{equation}
i(\Omega - \omega) v_{\phi}^{\prime} + \frac{\kappa^{2}}{2\Omega} v_{r}^{\prime} = - \frac{ih^{\prime}}{r},
\end{equation}
\begin{equation}
i(\Omega - \omega) v_{z}^{\prime} = - \frac{\partial h^{\prime}}{\partial z},
\end{equation}
\begin{eqnarray}
i(\Omega - \omega) h^{\prime} & +& v_{r}^{\prime} \frac{\partial h}{\partial r} + v_{z}^{\prime} \frac{\partial h}{\partial z}  \nonumber \\
           &  & = - c_{s}^{2}  \left[ \frac{1}{r} \frac{\partial (r v_{r}^{\prime})}{\partial r}
           + \frac{iv_{\phi}^{\prime}}{r} + \frac{\partial v_{z}^{\prime}}{\partial z} \right],
\end{eqnarray}
where $\kappa = \left[ 2 \Omega \left( 2\Omega + r d \Omega/dr \right) \right]^{1/2}$
is the epicyclic frequency
\footnote{In our model, the detailed form of $\kappa$ is given by
\begin{eqnarray}
\kappa & = & \left( \frac{GM_{1}}{r^{3}} \right)^{1/2}
           \left\{ 1 - 2q \left(\frac{r}{D} \right)^{3}
            - k_{2} \left( \frac{\Omega_{1}}{\Omega_{\rm c}} \right)^{2}
            \left( \frac{R_{1}}{r} \right)^{2} \right. \nonumber\\
      && \left. + \left[ 2 \left( \frac{d \ln \rho_{m}}{ d \ln r} \right)
         + \frac{d^{2} \ln \rho_{m}}{d \ln r^{2}} \right]
         \left( \frac{H}{r} \right)^{2} \right\}^{1/2}.
      \nonumber
\end{eqnarray}
},
($v_{r}^{\prime}, v_{\phi}^{\prime}, v_{z}^{\prime}$) is the perturbed velocity, and
$h^{\prime}$ is the enthalpy perturbation, which is related to the density perturbation
$\rho^{\prime}$ via $h^{\prime} = c_{\rm s}^2 \rho^{\prime}/\rho$.

In order to take the three-dimensional effect into account, we expand perturbed quantities in the $z$-direction in terms of Hermite polynomials as
\begin{equation}
v_{r}^{\prime} (r,z) = \sum_{n} u_{n}(r)H_{n}(\zeta),
\label{eq:vrprime}
\end{equation}
\begin{equation}
v_{\phi }^{\prime} (r,z) = \sum_{n} v_{n}(r)H_{n}(\zeta),
\label{eq:vphiprime}
\end{equation}
\begin{equation}
v_{z}^{\prime} (r,z) = \sum_{n} w_{n}(r)H_{n-1}(\zeta),
\label{eq:vzprime}
\end{equation}
\begin{equation}
h^{\prime} (r,z) = \sum_{n} h_{n}(r)H_{n}(\zeta),
\label{eq:hprime}
\end{equation}
where $\mathit{H}_n(\zeta)$ is the Hermite polynomial defined by
\begin{equation}
H_{n}(\zeta) = \exp \left(\frac{\zeta^{2}}{2} \right)
   \left(-\frac{d}{d \zeta} \right)^{n} \exp \left(-\frac{\zeta^{2}}{2} \right)
\end{equation}
with $\zeta = z/H$ and $n$ being a dimensionless vertical coordinate
and a non-negative integer, respectively \citep{Ogi08}.

Following \citet{Ogi08}, we close the system of resulting equations by
assuming $u_{n}\equiv 0$ for $n\geq 2$.
Neglecting $u_{2}$ compared to $u_{0}$ is equivalent to
assuming the eccentricity of perturbed orbit of each gas particle to be independent of $z$.
After some manipulations, we have the following set of
equations to be solved for linear $m=1$ isothermal perturbations:

\begin{eqnarray}
\frac{du_{0}}{dr} &=& -\left(\frac{1}{2} + \frac{d\ln \Sigma}{d\ln r} \right) \frac{u_{0}}{r}
   - i \frac{\Omega}{c_\mathrm{s}^{2}} h_{0}
\label{eq:basic-1}
\\
i\frac{dh_{0}}{dr} &=& -2\left[\omega-\left(\omega_\mathrm{pr}
   + \frac{9c_\mathrm{s}^{2}}{4r^{2}\Omega}\right)  \right] u_{0} - 2i \frac{h_{0}}{r},
\label{eq:basic-2}
\end{eqnarray}
and
\begin{equation}
ih_{2} = \frac{3 c_\mathrm{s}^{2}}{2r\Omega} u_{0},
\label{eq:A8}
\end{equation}
where $\omega_\mathrm{pr}(r) \equiv \Omega(r) - \kappa(r)$ is the local apsidal precession frequency.
In deriving these equations, we have used approximations
$|\omega| \ll \Omega$, $|\Omega-\kappa| \ll \Omega$, and $(c_{\rm s}/r\Omega)^{2} \ll 1$.

Let us now consider the boundary conditions for equations~(\ref{eq:basic-1}) and (\ref{eq:basic-2}).
In the disk formation stage, when material is injected continuously from the star,
there is no gap between the star and the disk. Since the pressure scale-height of the Be star near the interface is much smaller than that of the disk, the oscillations cannot penetrate the stellar surface. Hence, we take the rigid wall boundary condition at the disk inner radius,
i.e., $u_{0} = 0$ at $r=R_{1}$.
On the other hand, in the disk dissipation stage, a gap forms between the star and the disk.
As a result, there appears a special radius near the star, inside of which
no stable circular orbit exists, i.e., $\kappa^2<0$.
We take this radius of innermost stable circular orbit, $r=r_{\rm ISCO}$, as the inner radius of the disk,
and applies the free boundary condition at this radius. This condition is reduced to
\begin{equation}
i(\Omega - \omega ) h_{0} + \frac{d\ln\Sigma}{d\ln r} \frac{c_\mathrm{s}^{2}}{r} u_{0}=0
\end{equation}
\citep{Oct09} at $r=r_{\rm ISCO}$. Note that in general the innermost stable circular orbit appears
only in general relativistic disks, not in Newtonian disks.
In the current case, $\kappa^{2}$ becomes zero because of a large negative value
of $d^2\ln \rho_{\rm m}/d\ln r^2$ near the star due to gap formation.

In disks around isolated Be stars, eigenmodes must be confined to an inner region surrounded by a wide evanescent region where modes exponentially decay. Thus, as the outer boundary condition in the case of isolated Be stars, we take that the radial velocity perturbation $u_{0}$ vanishes
at a radius deep into the evanescent region.
For binary Be stars, we take the free boundary condition at $r=R_{\rm tides}$
as the outer boundary condition, because the disk is truncated at this radius.

\subsection{Eigenmodes for Isolated Be Stars}
We calculate the eigenfunctions as follows. First, we take the density distribution at one particular epoch, e.g., one year after the disk formation stage begins. Then, we solve equations~(\ref{eq:basic-1}) and ~(\ref{eq:basic-2}) with the boundary conditions mentioned above, using that density distribution. Once an eigenmode is found, the $z$-dependent part of the enthalpy perturbation is obtained from equations~(\ref{eq:hprime}) and (\ref{eq:A8}). We perform these calculations for several epochs. In this subsection we show the results for isolated Be stars.

\begin{figure}
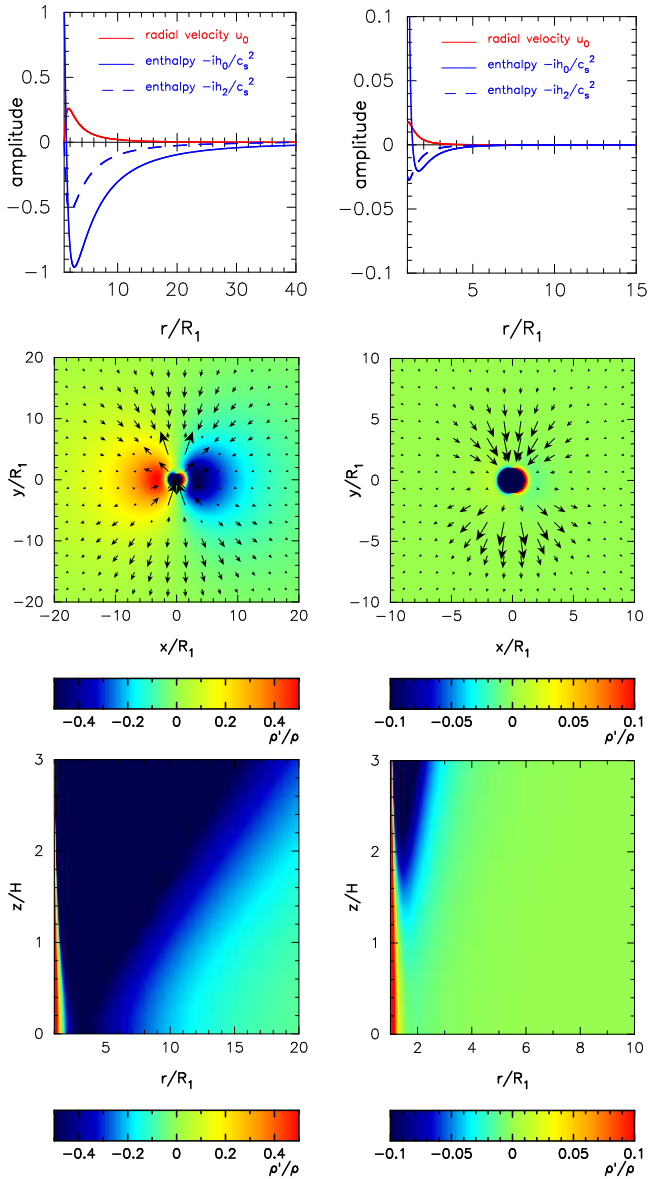

   \begin{center}
	     \begin{tabular}{cc}
	       \includegraphics[width=40mm]{ibygraffn10.ps} & \includegraphics[width=41mm]{ibygrafdn10.ps} \\
	       \includegraphics[width=40mm]{ibdmfn10.ps} & \includegraphics[width=40mm]{ibdmdn10.ps} \\
	       \includegraphics[width=40mm]{ibtrdfn10.ps} & \includegraphics[width=40mm]{ibtrddn10.ps}
		     \end{tabular}
	\end{center}		
		\caption{The $m=1$ fundamental mode in the formation (left) and dissipation (right) stages
		for an isolated Be star after 10 years of disk formation and dissipation,
		respectively.
		Top: The radial velocity perturbation (red line) and two components of enthalpy perturbation,
        $h_{0}$ (blue solid line) and $h_{2}$ (blue dashed line), as a function of radius.
        The density perturbation relative to the unperturbed density,
        $\rho^{\prime}/\rho$, is given by
        $\rho^{\prime}/\rho = \{h_{0}(r)+h_{2}(r)[(z/H)2-1)]\}/c_{\rm s}^2$.
		Middle: Perturbation pattern in the equatorial plane ($z=0$). The color-scale plot shows
		the density perturbation relative to the unperturbed density,
		$\rho^{\prime}/\rho$, while the arrows denote the perturbed velocity vectors.
		Bottom: The relative density perturbation, $\rho^{\prime}/\rho$, in the $(r,z)$-plane at
		$\phi-\omega t = 0$.}
    \label{fig:eigsing}
\end{figure}

Fig.~\ref{fig:eigsing} compares the $m=1$ fundamental modes in the formation stage (left panels) and dissipation stage (right panels) $10$ years after the disk formation/dissipation started. The top panels show the radial velocity perturbation $u_{0}$ (red line), the two components of enthalpy perturbations, $h_{0}$ (blue solid line) and $h_{2}$ (blue dashed line), as a function of radius.  Note that the density perturbation relative to the unperturbed density, $\rho^{\prime}/\rho$, is given by $\rho^{\prime}/\rho = \{h_{0}(r)+h_{2}(r)[(z/H)2-1)]\}/c_{\rm s}^2$. The modes are linear, so that the normalization is arbitrary. We see that in the disk formation stage, the eigenfunction is confined to a region within $r \sim 30$ $R_{1}$. On the other hand, the confinement of the mode is much more prominent in the disk dissipation stage,
the fundamental mode being confined within $r \sim 5$ $ R_{1}$.

In the middle panels of Fig.\ref{fig:eigsing}, we show the density perturbation relative to the unperturbed density, $\rho^{\prime}/\rho$, in the $(r,\phi)$-plane. The red (blue) region has a positive (negative) density perturbation.
The arrows superposed on the color-scale plot denote the velocity vectors associated with the mode.
From a close look at the middle panels, we note that regions with positive (negative) density perturbation
have negative (positive) perturbation of the azimuthal velocity. This implies that
when the violet (red) peak is stronger than the red (violet) peak,
the line profile as a whole will shift to a redder (bluer) wavelength. This is a typical behavior
of the observed long-term line-profile variability.
The bottom panels present the relative density perturbation, $\rho^{\prime}/\rho$, in the $(r,z)$-plane,
at $\phi - \omega t = 0$. As found in \citet{Oct09}, the density perturbation in the vertical direction
does not always take the maximum in the disk midplane.

As seen in section \ref{sec:evol}, the density distribution of an isothermal decretion disk
in the formation stage approaches asymptotically to the $\Sigma \sim r^{-2}$ distribution.
Since the $m=1$ eigenmode depends on the disk structure,
the eigenfrequency in the disk formation stage also approaches asymptotically to
that for the $\Sigma \sim r^{-2}$ distribution. This is shown in Fig.~\ref{fig:isbe},
where the frequency of the fundamental $m=1$ mode in the disk formation stage
is denoted by the thick solid line. Figure~\ref{fig:isbe} also shows that
the fundamental mode 
in the dissipation stage (thin solid line) has significantly higher eigenfrequency than
in the formation stage: When the mass injection stops and a gap starts opening,
the eigenfrequency jumps up by about a factor of three.
This sudden increase of eigenfrequency is due to the increase of
local precession frequency $\omega_{pr}=\Omega-\kappa$, which is caused by
the rapid decrease of epicyclic frequency from $\kappa \sim \Omega$ to $\kappa \sim 0$ near the star.
Afterwards, the eigenfrequency almost stays constant throughout the dissipation stage.
Note that only prograde modes, i.e., modes with positive frequencies
{(\citealt{Sav93})}, can exist in disks around isolated Be stars.

\begin{figure}
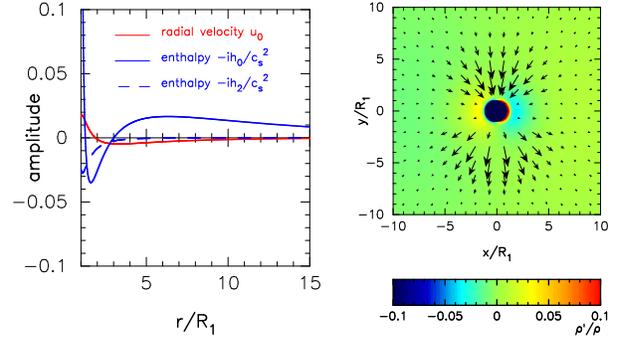


    \begin{tabular}{cc}
	    \includegraphics[width=41mm]{ibygrafdn10x.ps} & \includegraphics[width=34mm]{ibdmdn10x.ps}
		\end{tabular}

     \caption{The $m=1$ first overtone in the dissipation stage
     for an isolated Be star after 10 years of disk dissipation.
     Left: The radial velocity perturbation (red line) and the two components of enthalpy perturbation,
        $h_{0}$ (blue solid line) and $h_{2}$ (blue dashed line), as a function of radius.
     Right: Perturbation pattern in the $(r,\phi)$-plane. The color-scale plot shows
     the density perturbation normalized by the unperturbed density,
     $\rho^{\prime}/\rho$, while the arrows denote the perturbed velocity vectors.}
    \label{fig:eigsingfm}
\end{figure}

\begin{figure}
	\begin{center}
     \includegraphics[width=60mm]{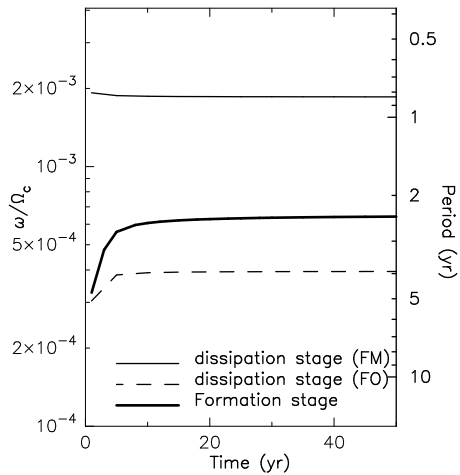} 	
	\end{center}
	\caption{Evolution of the eigenfrequency for an isolated Be star. Thick solid line denotes the eigenfrequency of the fundamental mode in the disk formation stage, while the thin solid line denotes that in the disk dissipation stage. The eigenfrequency of the first overtone in the disk dissipation stage
	is also shown by the thin solid line.}
    \label{fig:isbe}
\end{figure}

As mentioned above, the fundamental eigenmode in the dissipation stage has a much narrower
propagation region than in the formation stage.{
Although probing such a mode may be possible by detailed analysis of weak lines arising from the innermost disk region, it is is still challenging compared to the observation of modes in the disk formation stage, given the relatively small contribution to the total disk emission.}
Therefore, it is interesting to see that first overtones exist
in the disk dissipation stage, which have comparable size of propagation region
(and eigenfrequency) to that of fundamental modes in the disk formation stage
(Fig.~\ref{fig:eigsingfm}).
These modes could show up in observed line profiles when Be disks are dissipating.
Note that only fundamental modes exist in forming disks around isolated Be stars.

\subsection{Effect of the Companion Star}
Using the surface density distribution obtained in section~\ref{sec:evol-bin},
we have studied $m=1$ eigenmodes in Be disks in binary systems.

\begin{figure}
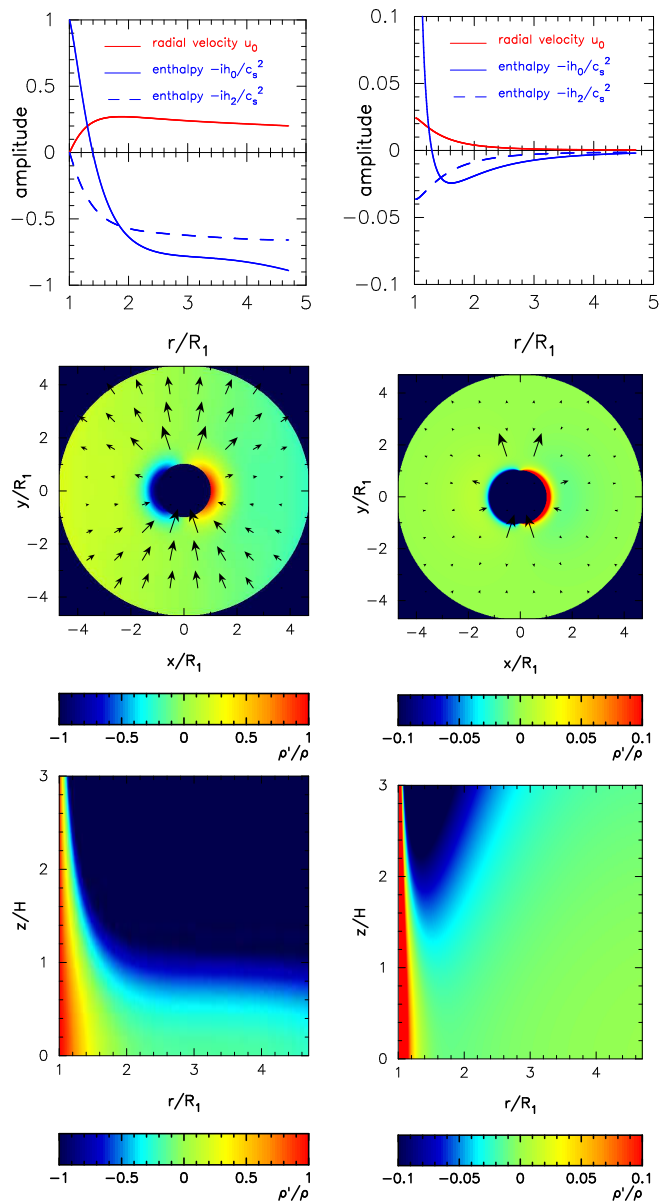

  \begin{center}
    \begin{tabular}{cc}
	       \includegraphics[width=40mm]{bbygraffd1n10.ps} & \includegraphics[width=42mm]{bbygrafdd1n10.ps} \\
	       \includegraphics[width=40mm]{bbdmfd1n10.ps} &  \includegraphics[width=40mm]{bbdmdd1n10.ps} \\
	      \includegraphics[width=40mm]{bbtrdfd1n10.ps} &   \includegraphics[width=40mm]{bbtrddd1n10.ps}
		\end{tabular}
   \end{center}	
			\caption{The $m=1$ fundamental mode in the formation (left) and dissipation (right) stages
        for a Be binary with $D = 10 R_{1}$ and $q=0.1$ after 10 years of
        disk formation and dissipation, respectively.
        The format of the figure is the same as that of Fig.~\ref{fig:eigsing}.}
    \label{fig:eigbin}
\end{figure}

\begin{figure}
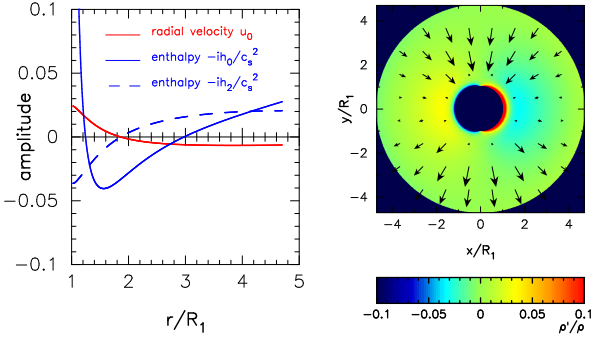

	\begin{center}
         \begin{tabular}{cc}
	       \includegraphics[width=40mm]{bbygrafdd1n10x.ps} & 	\includegraphics[width=34mm]{bbdmdd1n10x.ps}
		\end{tabular}
	\end{center}	
     \caption{The $m=1$ first overtone in the dissipation stage
     for a binary Be star with $D = 10 R_{1}$ and $q = 0.1$.
     The format of the figure is the same as that of Fig.~\ref{fig:eigsingfm}.}
    \label{fig:eigbinfo}
\end{figure}

\begin{figure}
  \begin{center}
    \begin{tabular}{cc}
	       \includegraphics[width=40mm]{bbygraffd3n12.ps} & \includegraphics[width=40mm]{bbygrafdd3n12.ps} \\
	       \includegraphics[width=40mm]{bbdmfd3n12.ps} &  \includegraphics[width=40mm]{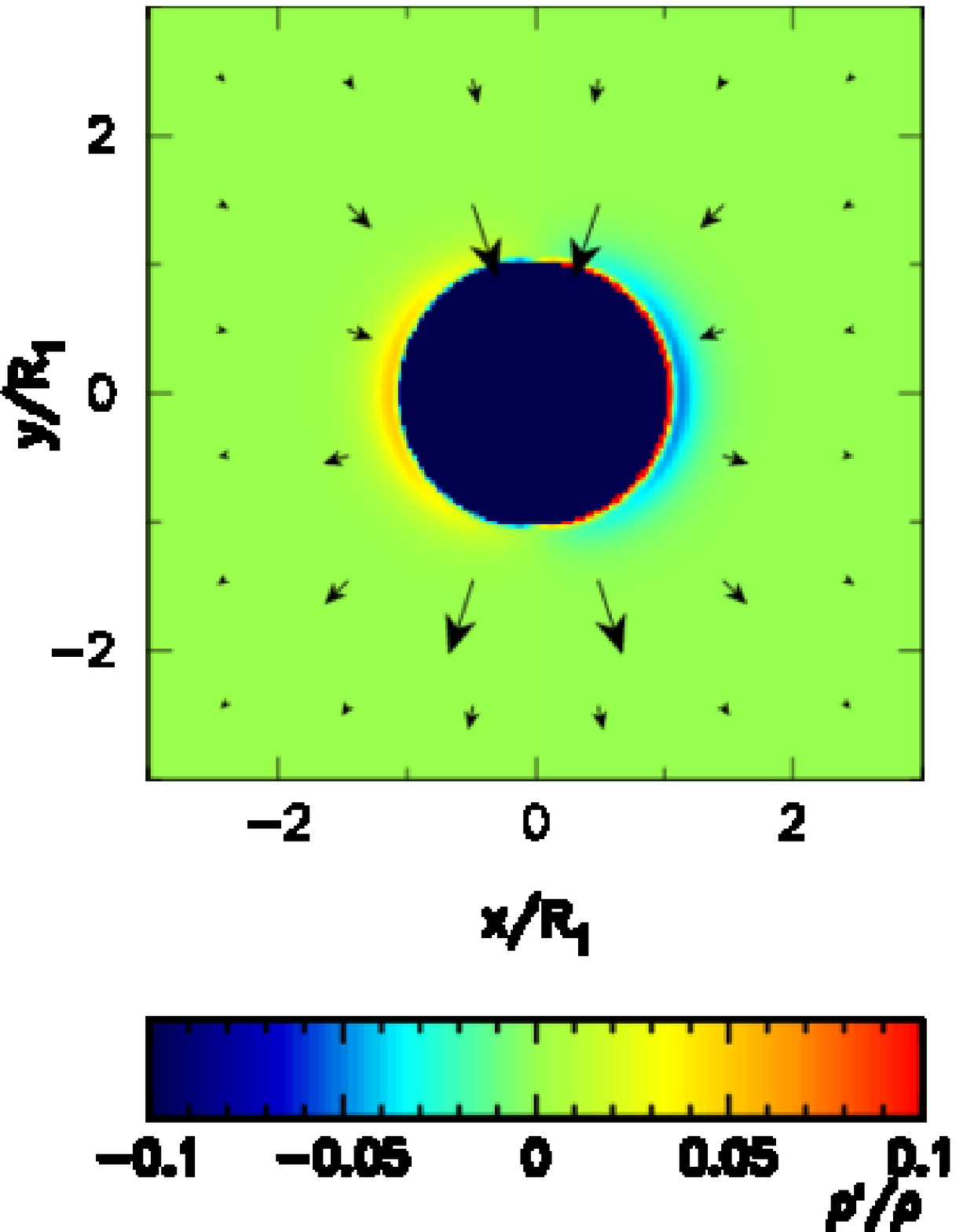} \\
	      \includegraphics[width=40mm]{bbtrdfd3n12.ps} &   \includegraphics[width=40mm]{bbtrddd3n12.ps}
		\end{tabular}
   \end{center}	
			\caption{The $m=1$ fundamental mode in the formation (left) and dissipation (right) stages
        for a Be binary with $D = 30 R_{1}$ and $q=0.1$ after 10 years of
        disk formation and dissipation, respectively.
        The format of the figure is the same as that of Fig.~\ref{fig:eigsing}.}
    \label{fig:eigbin3}
\end{figure}

Figure ~\ref{fig:eigbin} shows the fundamental $m=1$ mode in the formation (left panels) and
dissipation (right panels) stages for a binary with orbital separation $D = 10 R_{1}$ and
binary mass ratio $q = 0.1$ after 10 years of disk formation and
dissipation, respectively. The {format} of the figures is the same as
{that of} Fig.~\ref{fig:eigsing}. {As seen from the figure, the
  fundamental mode for a binary system with $D = 10$ $R_{1}$ propagates
  over the whole disk. However, for much wider binary systems (e.g.,
  those with $D \geq 25$ $R_{1}$), we have found eigenmodes that are
  confined to the { inner part} of the disk {as seen in Figure ~\ref{fig:eigbin3}}. This is because in wide binary systems the disk is large enough to confine these modes. We found that the correlation between radial velocity and density perturbations is consistent with observed line-profile variability, as in the isolated Be star case. We also have similar vertical density distribution to those found for isolated Be star. We show the first overtone found in the dissipation stage in Fig.~\ref{fig:eigbinfo} {for a binary system with $D = 10$ $R_{1}$, }since the fundamental mode found in this stage also propagates in a narrow region that may be {challenging} to observe.

\begin{figure}
	\begin{center}
     \includegraphics[width=60mm]{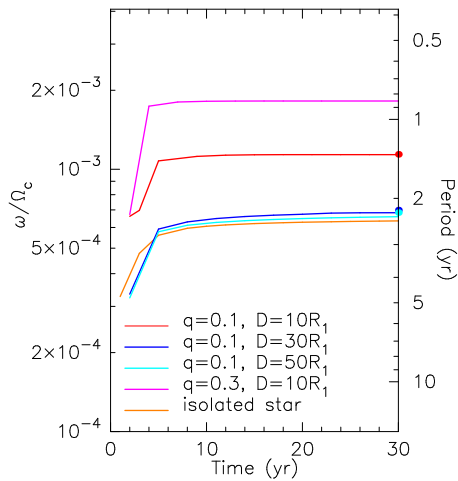} 	
	\end{center}			
	\caption{{Evolution of the eigenfrequency in the disk formation stage for binaries and an isolated Be star. The lines denote the eigenfrequency of the fundamental mode for binary systems with $q=0.1 $ and binary separation $D=10\,R_{1}$, $30\,R_{1}$, and $50\,R_{1}$ to observe the effect of binary separation. To observe the effect of binary mass ratio, we plot the results for a binary system with  $q=0.1$ and $D=10\,R_{1}$. 
	For comparison, the {dots} at $t = 30\,\rm{yr}$ denote the eigenfrequencies from the calculation with the $\Sigma \sim r^{-2}$ density distribution \citep{Oct09}.}}
    \label{fig:perbf}
\end{figure}

\begin{figure}
	\begin{center}
      \includegraphics[width=60mm]{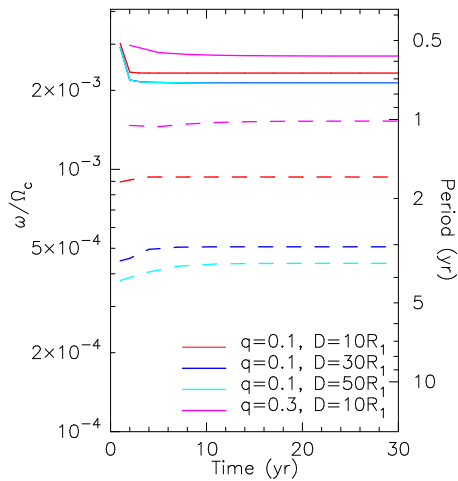}
	\end{center}					
	\caption{{Evolution of the eigenfrequency in the disk dissipation stage for binary Be stars.  The solid lines are for the fundamental modes and the dashed lines are for the first overtones.}
	 }
    \label{fig:perbd}
\end{figure}

The effect of binary separation and {binary mass ratio} on the global one-armed oscillations in the disk formation stage is shown in Fig.~\ref{fig:perbf}. The {red, blue, and cyan} lines denote the eigenfrequency of the fundamental mode for binary systems with $q=0.1$ and $D=10\,R_{1}$, $30\,R_{1}$, and $50\,R_{1}$, respectively. {The magenta line denotes the eigenfrequency of the fundamental mode for a binary system with $q=0.3$ and $D=10\,R_{1}$.} For comparison, we also show the eigenfrequency for the isolated Be star case by the {orange} lines, and those for steady disks with $\Sigma \propto r^{-2}$ by {dots} \citep{Oct09}. As in the case of isolated Be stars, the eigenfrequency gradually increases with time,
approaching that for the steady disk case. We also see that the eigenfrequency is higher for a closer binary {and a binary with a higher mass ratio. Note that the eigenfrequencies for $D=30\,R_{1}$ and $50\,R_{1}$ are almost indistinguishable, because the mode is already well confined for $D=30\,R_{1}$.}

Fig.~\ref{fig:perbd} shows the eigenfrequency in the disk dissipation stage. {Solid} lines are for the fundamental modes, while {dashed} lines are for the first overtones. As in Fig.~\ref{fig:perbf},
the {red, blue, cyan, and magenta} lines denote the eigenfrequency for binary {systems with $q=0.1$ and $D=10\,R_{1}$, $30\,R_{1}$, and $50\,R_{1}$, and a binary system with $q=0.3$ and $D=10\,R_{1}$  respectively.} As in the case of isolated Be stars, the eigenfrequency
in the disk dissipation stage in binary Be stars is also significantly higher than
that in the disk formation stage. It changes little throughout the dissipation stage,
except for very early epochs.
The dependence of the eigenfrequency on the orbital separation {and binary mass ratio} in the disk dissipation stage is
the same as in the disk formation stage. 
{Note also that the eigenfrequencies of the fundamental modes for $D=30\,R_{1}$ and $50\,R_{1}$ are now indistinguishable, because of the better mode confinement than in the formation stage.}

We have also studied the one-armed modes for binaries with $q=0.3$ in order to investigate the effect
of binary mass ratio on the modes. We found that the eigenfrequency is higher in systems
with a higher binary mass ratio, for both disk formation and dissipation stages.
This is because the local precession frequency, $\omega_{pr}$ increases with increasing mass ratio $q$.
For example, for a system with $D = 10 R_{1}$ and $q = 0.1$ in the
formation stage, we have $\omega= 1.13\times 10^{-3}$ { $\Omega_{c}$} at $t=30$ yr, while for the system with the same binary separation but
a higher binary mass ratio $q = 0.3$, we have $\omega= 1.81 \times
10^{-3}$ { $\Omega_{c}$} at the same epoch.

From the results above, the dependence of the eigenfrequency of the one-armed modes
in evolving Be disks on the orbital separation and binary mass ratio is qualitatively
the same as in steady ($\Sigma \sim r^{-2}$) disks around binary Be stars.

\section{Discussion}
\label{sec:discussion}

We first compare the $m=1$ eigenfrequencies found in the formation
stage of Be disks in binary systems with those calculated using a power-law density distribution (\citealt{Oct09}). As discussed above, we
found a similar dependence of the eigenfrequency on the binary
parameters such as orbital separation and binary mass ratio: The
eigenfrequency decreases with increasing binary separation and/or
decreasing binary mass ratio. Hence, the oscillation period is
shorter for close binary systems than for wide binary
systems. Moreover, the eigenfrequencies at later times in the
disk formation stage are closer to those for steady disks than the
  eigenfrequencies at earlier times in this stage. This is
expected because, from the evolution of the surface density
distribution shown in Fig{s}.~\ref{fig:devob} and ~\ref{fig:devob5} ,
we can see that the surface density distribution in the latter part
of the disk formation stage is almost similar to that of the steady
disk case where $\Sigma \sim r^{-2}$, especially in the inner part of the disk.

Another interesting result is on the surface density distribution of binary Be disks in the disk dissipation stage. In our results, the surface density distribution in the inner part of the disk decreases more rapidly in close binary systems than in wide binary systems. Observationally, there is an indication that the disk is lost from the innermost part (e.g., \citealt{Riv01}, \citealt{WI10}). However, we do not know what mechanism causes this gap. It can be the result of the accretion of the material in the inner part of the disk due to viscosity once the mass injection from the star stops, or the ablation of the disk by radiation  from the central star can also drive away the material in the innermost part, which is closest to the star, to the outer part of disk. Our results show that if the cause of the gap in the disk dissipation stage is the accretion due to viscosity, then we can observe difference in the decreasing rate of surface density between the close and wide binary systems. If there are no difference then the gap in the inner-part of the disk should be mainly caused by ablation by the radiation of the central star.

Recently, {\citet{Car12}} compared the observed visual lightcurve {in the disk dissipation phase} of the
Be star $28$ CMa with lightcurves obtained from dynamical viscous
decretion disk model with different values of viscosity parameter,
$\alpha$. Interestingly, the result from the model with {$\alpha = 1.0$}
is the one that reproduces best the observed lightcurve in all
phases. Hitherto, most of the theoretical studies of Be disks adopted
viscosity parameter $\alpha \sim 0.1$.  This value of $\alpha$ is thought
to be 
plausible, because studies of the formation and dissipation of Be disks (e.g., \citealt{CL01}, \citealt{WI10}) found that the timescale of disk formation/dissipation agrees with the viscous timescale for $\alpha \sim 0.1$ {and because the turbulent viscosity excited by the magneto-rotational instability in accretion disks is saturated around this level}. However, the result shown by {\citet{Car12} suggests that study for a much higher viscosity in disks of Be stars is needed}.

The effect of viscosity on $m=1$ modes in Be disks can be seen, for
example, in \citet{Car09} and \citet{Oka00}. When the effect of
viscosity is included, the mode becomes overstable and has a
one-armed, spiral density-perturbation pattern, the confinement of
which is weaker than neutral modes in inviscid disks. The one
-armed
spiral is more remarkable for a higher viscosity parameter. However,
for the viscosity parameter $\alpha \la 0.1$, the $m=1$ mode does not differ much from the inviscid case.  Moreover, the eigenfrequency is insensitive to $\alpha$, according to \citet{Oka00}. Thus, although in this paper we did not include viscosity in our eigenmode calculations, we believe that our results, particularly eigenfrequencies, are robust. However, our model cannot be applied to higher $\alpha$. Therefore, the next step will be studying $m=1$ eigenmodes by consistently taking into account the viscosity effect.

Below, we discuss the evolution of {$m=1$} eigenmodes in Be disk{s}. In the disk formation stage, the oscillation period decreases as the disk grows. However, the change  becomes small once the disk is fully developed (e.g., several years after the disk formation begins). Observationally, this small change in the oscillation period will difficult to be detected. Hence, we do not expect a significant change in the V/R variation period in Be disks at later times of the disk formation stage.

The disk dissipation stage starts right after the mass ejection from the star stops. One of the interesting results in this stage is that the fundamental mode has a significantly higher eigenfrequency and a much narrower propagation region.  Hence, the eigenmodes in forming/steady disks will disappear once the ejection of matter from the star stops and accretion starts in the inner part of the disk. The disappearance of eigenmodes in the disk dissipation stage is consistent to the observation of V/R variability on Be/X-ray binary LS V +44 17 by \citet{Rei05}. They show that the V/R variability disappeared before the complete loss of the disk.
They also argue that the effects of the density perturbation do not manifest themselves until the disk is fully developed.
This could be due either to slow excitation of the $m=1$ mode or
to the rapidly changing eigenstate in a forming disk.

In the dissipating disk, the fundamental $m=1$ mode is likely difficult to
observe due to the narrowness of the wave propagation region.
However, we may observe the first overtone if the mode {is} excited. In our model, the first overtone in the disk dissipation stage has a propagation region comparable to that of the fundamental mode in the formation stage. Thus, an $m=1$ eigenmode in a Be disk starts to decay when the disk starts accreting, and if the disk dissipation stage is long enough, another mode with different oscillation period and structure may appear.

{In our model, we use a simple process where mass is injected to the disk at a constant rate during the formation stage, and the mass injection {is shut off} during the dissipation stage. However, there are various scenarios for the mass injection to the disk such as periodic mass injection where the mass injection rate varies periodically, or episodic mass injection where there is sudden increase of the disk density caused by an outburst of mass injection. \citet{HA12} studied different scenarios for the mass injection mechanism to the Be disk and found different density distribution for disk formation stage and dissipation stage. As we can see in our result, the disk density distribution affect the one-armed oscillation modes. Therefore, it is interesting to study different mass injection scenarios for further studies of the one-armed oscillation modes with a more realistic disk model. }

Quantitatively, periods of $m=1$ eigenmodes obtained from our model do not agree with the observed V/R periods. The period of {$m=1$ modes} for both isolated and binary Be stars studied in this paper is in the range of a few years. For example, \citet{Rei05} listed V/R variation periods for some Be/X-ray binaries, binary systems that consist of a Be star and a neutron star, which is about a few years. However, the observed V/R variation period for Be stars in other class of binary systems is sometimes much longer than those of the Be/X-ray binaries. \citet{Ste07} show that the quasi-periodic V/R variation period in some binary Be stars ranges between 5-10 years. The situation is worse for isolated Be stars: The observed V/R variation period for isolated Be stars, which is typically 5-10 years (see \citealt{Oka97}), is much longer than the period obtained from the current model. The difference between the observed and model V/R variation periods implies that there are still missing mechanisms other than the radial density distribution that affect the global disk oscillations. As mentioned in \citet{Oct09}, taking into account a more realistic temperature distribution in the disk and/or the effect of optically thick line forces may contribute to lower the eigenfrequencies.

\section{Conclusions}
\label{sec:conclusion}

We have studied the effect of density distribution evolution on the global oscillation modes in equatorial disks around isolated and binary Be stars. We modeled the surface density distribution in the disk formation stage by injecting mass at a radius just outside the star, while for modeling the disk dissipation stage, we turned off the mass injection. We found that the disk density distribution is different from the usually assumed, single power-law density distribution, especially in the disk dissipation stage, when the inner part of the disk accretes, forming a gap between the star and the disk. For the global one-armed oscillations in the disk {formation} stage, we found that the eigenfrequencies gradually increase with time toward the values for steady disks. On the other hand, the eigenfrequencies stay almost constant in the disk dissipation stage, which are significantly higher than those in the disk formation stage. By this frequency gap, eigenmodes in a forming/steady disk will disappear when the disk starts dissipating. More detailed theoretical study and observations of V/R variations in forming and dissipating Be disks are desirable to improve/test the results obtained in this paper.

\bigskip

We thank Masayuki Fujimoto and Stanley P. Owocki for helpful discussions. FO, CK, and Ap acknowledges Research and Innovation 2015 Grant from ITB. This work was also partially supported by the Grant-in-Aid for Scientific Research (20540236) {and a research grand from Hokkai-Gakuen Educational Foundation}.

\label{lastpage}

\end{document}